\documentclass[fleqn,10pt]{wlscirep}
\usepackage[utf8]{inputenc}
\usepackage[T1]{fontenc}
\usepackage{array,multirow}
\usepackage{graphicx}
\usepackage{caption}
\usepackage{subcaption}
\usepackage{makecell}
\usepackage{xurl}
\usepackage{xcolor}
\usepackage{easyReview}
\usepackage{ulem}
\usepackage{lineno}
\title{Emphysema Subtyping on Thoracic Computed Tomography Scans using Deep Neural Networks}

\author[1]{Weiyi Xie}
\author[1]{Colin Jacobs}
\author[2]{Jean-Paul Charbonnier}
\author[3]{Dirk Jan Slebos}
\author[1,*]{Bram van Ginneken}

\affil[1]{Radboud University Medical Center, Department of Medical Imaging, Nijmegen, The Netherlands}
\affil[2]{Thirona, 6525 EC Nijmegen, The Netherlands.}
\affil[3]{University of Groningen, University Medical Center Groningen, Department of Pulmonary Diseases, Groningen, The Netherlands}
\affil[*]{bram.vanginneken@radboudumc.nl}

\begin{abstract}
Accurate identification of emphysema subtypes and severity is crucial for effective management of COPD and the study of disease heterogeneity. Manual analysis of emphysema subtypes and severity is laborious and subjective. To address this challenge, we present a deep learning-based approach for automating the Fleischner Society's visual score system for emphysema subtyping and severity analysis. 
We trained and evaluated our algorithm using 9650 subjects from the COPDGene study. Our algorithm achieved the predictive accuracy at 52\%, outperforming a previously published method's accuracy of 45\%. In addition, the agreement between the predicted scores of our method and the visual scores was good, where the previous method obtained only moderate agreement. Our approach employs a regression training strategy to generate categorical labels while simultaneously producing high-resolution localized activation maps for visualizing the network predictions. By leveraging these dense activation maps, our method possesses the capability to compute the percentage of emphysema involvement per lung in addition to categorical severity scores. Furthermore, the proposed method extends its predictive capabilities beyond centrilobular emphysema to include paraseptal emphysema subtypes.
\end{abstract}
\begin{document}

\maketitle
\thispagestyle{empty}

\section*{Introduction}
Chronic obstructive pulmonary disease (COPD) is the third leading cause of death worldwide, causing 3.23 million deaths in 2019 \cite{WHO20}. COPD is characterized by irreversible airway obstruction, which can be caused by small airway diseases and emphysema. However, the extent to which each contributes to airflow obstruction varies among individuals, leading to significant heterogeneity among COPD patients \cite{Han10}. To better understand this heterogeneity, researchers have used computed tomography (CT) scans to assess the distribution, severity, and progression of the disease in vivo \cite{Coxs99,Mada06}. Additionally, many studies have attempted to identify COPD subtypes based on patterns observed on CT scans \cite{Smit14,Cast20,Yilm13,Lync15}.

The Fleischner Society has developed a structured scoring system for classifying subtypes of centrilobular and paraseptal emphysema based on the severity of the disease as observed on chest CT scans \cite{Lync15}. This scoring system, which we refer to in this work as the Fleischner system, utilizes six ordinal scales to evaluate centrilobular emphysema as absent, trace, mild, moderate, confluent, or advanced destructive, and paraseptal emphysema as absent, mild, or substantial. Previous studies of the Fleischner system \cite{Lync18,Hump20} have primarily focused on centrilobular emphysema and have reported good reader agreement in scoring centrilobular emphysema among human readers on scans from 3171 participants in the COPDGene study cohort \cite{Rega10} (a subset of the data used in our current study). These studies also demonstrated that visual scores of centrilobular emphysema are associated with mortality. To improve the efficiency of using the Fleischner system in research and clinical practice, a deep learning algorithm \cite{Hump20}, referred to in this work as the Humphries algorithm, has been developed to automatically score centrilobular emphysema according to the Fleischner system. When comparing the predicted scores of the Humphries algorithm with visual scores for 7143 subjects from the COPDGene study, the Humphries algorithm achieved a linear weighted kappa statistic of 0.60 and a classification accuracy of 45\%. Furthermore, the scores produced by the Humphries algorithm were found to be associated with mortality \cite{Hump20}.

In this paper, we present a novel method that addresses two limitations of the existing Humphries algorithm: its focus on centrilobular scores and its lack of model interpretability. Our proposed method not only calculates severity scores for centrilobular emphysema, but also for paraseptal emphysema, enabling the analysis of both subtypes. Furthermore, it generates more interpretable high-resolution emphysema activation maps, allowing for the quantification of the percentage of emphysema per lung for further analysis.

\section*{Methods}\label{sec:methods}
\subsection*{Data Collection and Partitioning}\label{sec:data}
In this study, we utilized chest CT scans from the COPDGene clinical trial, which included data from 21 imaging centers in the United States and enrolled 10,192 subjects between 2008 and 2011. From this dataset, we used a subset of 9650 subjects for analysis, which is the same data as used in the previous study by Humphries et al. \cite{Hump20}. Using the common data set (including the partitioning) enabling us to directly compare our algorithm with the Humphries's algorithm \cite{Hump20}. The cohort of 9650 subjects includes all available  COPDGene subjects having a baseline inspiration CT scan, the Fleischner visual scores, and mortality data during the Humphries's study. The lungs of these subjects on their inspiration CT scans were then segmented semi-automatically by trained analysts from Thirona, a company specializing in chest CT analysis.

Following the Humphries's study \cite{Hump20}, 9650 subjects were partitioned into a development set ($n=2507$) and an evaluation set ($n=7143$), with the development set further divided into a training set ($n=2407$) and a validation set ($n=100$). Only 2507 subjects were used as the development set because these were all available subjects with inspiration CT scans and Fleischner visual scores, excluding those previously included in an earlier analysis \cite{Lync18}. The evaluation set was used for statistical analysis and the reporting of performance metrics. The slice thickness of the CT scans ranged from 0.625-0.9mm and the pixel spacing ranged from 0.478-1.0mm. Most scans were performed using a tube voltage of 120kVp, a tube current of 200mAs, and reconstruction kernels B31f and B35f. The full CT protocols are detailed in Rega et al.\ \cite{Rega10}.

Table~\ref{tab:data} provides the distribution of COPD GOLD stages and Fleischner visual scores in the data selection, as well as the expected range of emphysema percentage per lung for each severity score. For example, a centrilobular emphysema score of 1 indicates an estimated emphysema percentage per lung between 1\% and 5\%. These percentage ranges were used to train our algorithm to capture the differences in disease severity scores. Patient demographics and lung functional parameters for the evaluation set ($n=7143$) can be found in the previous study by Humphries et al.\cite{Hump20}.

\begin{table}
  \caption{Distribution of GOLD stages \cite{Rega10} and the Fleischner visual scores \cite{Lync15} in our data selection in the development set (dev) and the evaluation set (eval). No PFT: spirometry data not available; PRISM: Preserved Ratio Impaired spirometry \cite{Wan14}.}
  \centering
  \begin{subtable}[t]{0.95\textwidth}
  \centering
  \begin{tabular}[t]{lll}
    &(a) GOLD stages  & \\
    \toprule
    GOLD stages     & \# subjects (dev)     & \#subjects (eval) \\
    \midrule
    GOLD0     & 981 &  3178     \\
    GOLD1     & 182 &  570     \\
    GOLD2     & 440 &  1371     \\
    GOLD3     & 305 &  771     \\
    GOLD4     & 205 &  337     \\
    Non PFT  &0 & 63 \\
    Non Smoking     & 70 &  36     \\
    PRISm     & 324 &  817     \\
    \bottomrule
     Total     & 2507 &  7143     \\
    \bottomrule
 \end{tabular}
\end{subtable}
 \hspace{\fill}
 \begin{subtable}[t]{0.45\textwidth}
 \begin{tabular}[t]{lll}
  &(b) Centrilobular scores & \\
  \toprule
  Score     & \makecell[l]{\#subjects \\ (dev)} 
  & \makecell[l]{\#subjects \\ (eval)}   \\
  \midrule
  0 (0-1\%)     & 782 &  2499     \\
  1 (1-5\%)    & 431 &  1322     \\
  2 (5-10\%)    & 478 &  1409     \\
  3 (10-20\%)    & 430 &  1049     \\
  4 (20-30\%)    & 275 &  656     \\
  5 (30-100\%)   & 111 &  208     \\
  \bottomrule
Total & 2507 & 7143\\
  \bottomrule
\end{tabular}
\end{subtable}
\begin{subtable}[t]{0.45\textwidth}
 \begin{tabular}[t]{lll}

  &(c) Paramseptal scores & \\
  \toprule
  Score     & \makecell[l]{\#subjects \\ (dev)} 
  & \makecell[l]{\#subjects \\ (eval)}  \\
  \midrule
  0 (0-1\%)     & 1145 &  3857     \\
  1 (1-5\%)    & 739 &  1865     \\
  2 (5-100\%)    & 623 &   1421    \\
  \bottomrule
Total & 2507 & 7143\\
  \bottomrule
\end{tabular}
\end{subtable}
 \label{tab:data}
\end{table}

\subsubsection*{Reference Standard}
CT scans were visually scored according to the Fleischner system by analysts who did not have prior experience in radiology interpretation. The annotation process is described in more detail in previous research by Lynch et al.\cite{Lync15,Lync18}. A reader study conducted on 3171 scans from our data selection by two analysts found good agreement in scoring centrilobular emphysema using the Fleischner system \cite{Lync18}.

\subsubsection*{Lung Segmentation}
The lungs were extracted using commercialized software (LungQ, Thirona, Nijmegen, NL), followed by manual refinement if needed. We used lung segmentations for preprocessing CT scans in developing our algorithm.  

\subsection*{Algorithm Design}
\subsubsection*{Pre-processing of CT scans}
All CT scans in this work were preprocessed by clamping the intensity values between $\left[-1150\sim-300\right]$, rescaling to $\left[0\sim1\right]$, cropping using lung segmentations, and resizing using trilinear interpolation to $128\times224\times288$ in the axial, coronal, and sagittal dimensions, respectively. Note that the spacing may not be isotropic following this resizing process. The primary goal of resizing the images to a fixed input size is to maintain consistent GPU memory consumption throughout the inference process, thereby enhancing the practical utility of the proposed algorithm. By utilizing an input size of $128\times224\times288$, we can accommodate a batch size of 4 within our network architecture while adhering to our computational budget (NVDIA A100 GPU with 40 gigabytes GPU memory). And the lung segmentations were used to set values outside of the lungs to zero, allowing the algorithm to focus on learning features within the lungs.

\subsubsection*{Neural Network Architecture}
\begin{figure*}
\centering
 \includegraphics[width=0.85\textwidth]{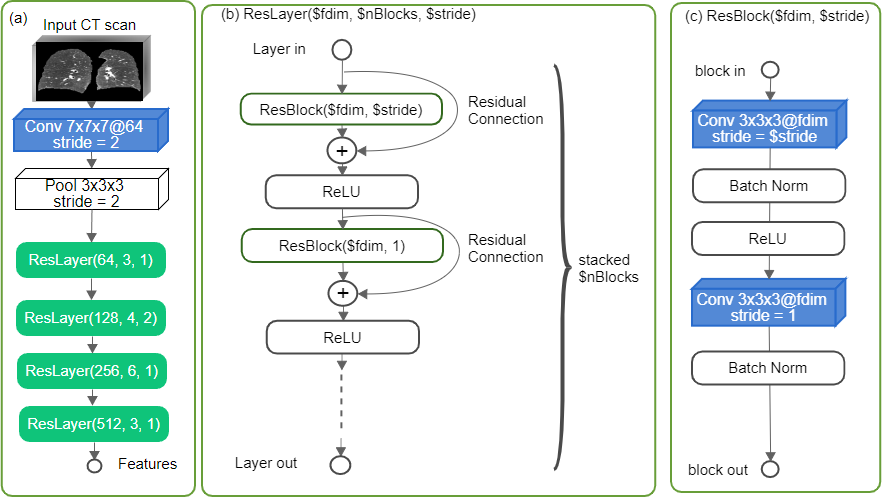}
\caption{The overview of our ResNet-based backbone with 34 layers (a), consisting of four stacked ResNet layers (b). Each layer consists of 3, 4, 6, and 3 ResNet blocks (c), respectively, from top to bottom. } 
\label{fig:framework}
\end{figure*}

In building our algorithm, we chose to use the ResNet variant with 34 layers \cite{He16} as the backbone, as ResNet is a widely used convolutional neural network architecture in image recognition. As shown in Fig.~\ref{fig:framework}, the network begins with a 3D convolution operation (kernel size=7, stride=2) using 64 filters to reduce the spatial size of the input by half. All convolutions in our network are isotropic, followed by batch normalization and a rectifier linear unit (ReLU) activation function, unless otherwise specified. We then reduce the spatial resolution by half using a max pooling layer (kernel size=3, stride=2). The pooled features serve as the input to four stacked ResNet layers, each consisting of several ResNet blocks. The number of ResNet blocks per layer increases from the first to the fourth layer, with 3, 4, 6, and 3 blocks, respectively. Each ResNet block includes two convolution operations with kernel size 3. From the second layer onwards, the first convolution in each block doubles the number of filters and, at the second layer, also reduces the spatial resolution by half using a stride of 2. This results in the input size being reduced by a factor of 8.

\begin{figure*}
\centering
 \includegraphics[width=0.85\textwidth]{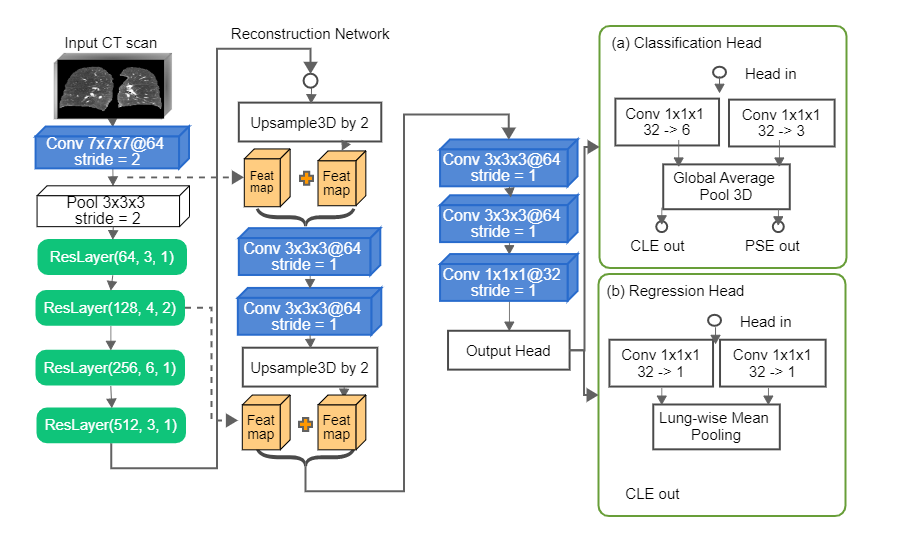}
\caption{The reconstruction network on top of the ResNet backbone for producing dense features. The output of the reconstruction network is fed into either classification or the regression output head depending on the training strategy for predicting centrilobular scores (CLE out) and paraseptal scores (PSE out). Note that we do not use both output heads together for multi-tasking. } 
\label{fig:reconstruction_net}
\end{figure*}

The Resnet backbone generates convolutional features at a resolution that is eight times lower than the input resolution, which can compromise the interpretability of the model using standard techniques such as class activation maps (as demonstrated in the Humphries algorithm \cite{Hump20}). To address this limitation, we propose the inclusion of a reconstruction network on top of the ResNet backbone for the generation of dense features. As shown in Figure~\ref{fig:reconstruction_net}, the reconstruction network takes features extracted from the Resnet backbone as the inputs and comprises two upsampling layers, each of which includes one trilinear upsampling operation and two 3x3 convolution filters to reduce upsampling artifacts. The upsampled features are then concatenated with the features from the ResNet backbone using a skip connection, as the one utilized in 3D-UNet \cite{Cice16}. The reconstruction network serves the same purpose as the decoder in 3D-UNet, where the ResNet backbone can be seen as the encoder.

On top of the reconstruction head, we utilize two output heads depending on the training strategy. 

For classification training, the dense features from the reconstruction network are reshaped to have the appropriate number of channels for the number of target classes. Two $1\times1\times1$ convolutions are employed to reshape the features, one for predicting centrilobular scores and the other for predicting paraseptal scores. The reshaped feature maps are then processed using global average pooling and activated using the softmax function to produce class probabilities. We refer to the reshaped dense features before pooling as dense class activation maps.

For regression training, the reconstructed features are reshaped into two single-channel feature maps using $1\times1\times1$ convolutions, one for predicting the percentage of centrilobular emphysema per lung and the other for predicting the percentage of paraseptal emphysema per lung. The sigmoid function is then applied to both feature maps, and the resulting features are averaged inside the lung segmentation to produce two floating numbers (emphysema percentages per lung). The sigmoid-activated dense features before pooling are referred to as the dense regression activation maps.

\subsubsection*{Algorithm Training}
We refer to the network with the classification output head as the classification network, and the network with the regression output head as the regression network.
We train each network separately and compare their results with the Humphries algorithm as the baseline. 

In the classification output head, we utilize a two-way convolutional layer with $1\times1\times1$ kernels to predict two targets, following the definitions of the Fleischner system. This is depicted in Figure~\ref{fig:reconstruction_net}a. The first target is to classify six levels of centrilobular severity: absent (0), trace (1), mild (2), moderate (3), confluent (4), or advanced destructive (5). The second target is to classify three levels of paraseptal severity: absent (0), mild (1), and substantial (2). During the training phase, we employ a weighted cross-entropy loss function, where the weights are initially determined by the inverse frequency of the target classes and are subsequently updated at the conclusion of each epoch in order to penalize classes with low per-class accuracy.

Training a convolutional neural network to classify emphysema severity scores in a multi-class classification setup may appear counter-intuitive. By definition, severity scores reflect the degree of involvement of emphysema in the lung and are directly correlated with the volume measurements of emphysematous regions. To correctly predict a severity score, the sum of the corresponding channel for that score in the class activation maps must be larger than the sum in the other channels, resulting in one channel suppressing activations in the other channels (being discriminative). As a result, the class activation maps generated by the trained network may not necessarily represent the underlying disease, emphysema, but rather highlight areas that were used to make the correct classification. For example, the zero-indexed channel corresponding to the "absent" class is expected to be more heavily activated than the other channels when there is no emphysema present in the image.

Therefore, instead of directly predicting visual scores, we propose a regression training strategy to estimate the percentage of emphysema involvement per lung using the interval regression loss proposed in previous research \cite{Xie21}. In the Fleischner system \cite{Lync15}, severity scores are defined based on the percentage of emphysema involvement in the lung. Therefore, we convert the severity scores into intervals of emphysema percentage per lung using a predefined mapping table. For example, a centrilobular severity score of 2 (mild) indicates that the estimated centrilobular emphysema-affected region should occupy between 1-5\% of the lung. In this way, each visual score corresponds to an interval of emphysema percentage in the lung. We use the mapping in Table~\ref{tab:data}b and Table~\ref{tab:data}c to convert visual scores into percentage intervals for training, and subsequently map the predicted percentage back to visual scores for evaluation. The regression output head produces two single-channel dense feature maps using a two-way convolution layer with $1\times1\times1$ kernels for estimating the emphysema percentage in the lung for centrilobular and paraseptal subtypes, respectively. Denoting the estimated percentage as $p$, and the target emphysema percentage range $(r_{l}, r_{u})$, the interval regression loss $L_{INT}$ can be written as:

\begin{equation}\label{eq:interval_regression_loss}
\begin{array}{l}
\qquad K = (0.5*(r_{l} - r_{u}))^2. \\
\qquad \mbox{min:~} L_{INT} = max(0, (p - 0.5*(r_{l} + r_{u}))^2-K ) \\
\end{array}
\end{equation} 
The dense regression activation maps, which are sigmoid-activated single-channel dense feature maps before the lung-wise mean pooling, represent the probability of emphysema. During training, we impose two constraints on the dense regression activation maps. The first constraint is the overlapping loss, which ensures that the dense regression activation maps for two emphysema subtypes do not overlap, as each voxel can only be assigned to one subtype (either centrilobular or paraseptal). Denoting the dense regression activation map for centrilobular emphysema as $p_{C}^{i}$ and the dense regression activation map for paraseptal emphysema as $p_{S}^{i}$ where $i$ is the location index, the overlapping loss $L_{OL}$ can be written as:
\begin{equation}\label{eq:overlapping_loss}
\begin{array}{l}
   \mbox{min:~} L_{OL} = \frac{2 * \sum_{i}(p_{C}^{i}  p_{S}^{i})}{\sum_{i}(p_{C}^{i}) + \sum_{i}(p_{S}^{i})}.
\end{array}
\end{equation} 
The overlapping loss is used to compute the soft Dice coefficient between two dense regression activation maps (probability maps). 

The second constraint is the segmentation loss. To generate the segmentation pseudo labels, we identify the low-attenuation areas (LAA-950) in the lungs by applying an intensity threshold of -950 HU to CT images. This threshold is commonly used in the analysis of emphysema on CT scans \cite{Lync15}. The segmentation loss ensures that the union of dense regression activation maps from both subtypes matches the LAA-950. Using the previous notations, the segmentation loss $L_{SEG}$ can be written as:
\begin{equation}\label{eq:segmentation_loss}
\begin{array}{l}
   p^{i} = min(max(0.0, p_{C}^{i} + p_{S}^{i}), 1.0) \\
   \mbox{min:~} L_{SEG} = \sum_{i}(-s*log (p^{i}*t^{i} + (1.0 - p^{i}) (1.0 - t^{i})))
\end{array}
\end{equation} 
where $t^{i}$ is the segmentation target indexed at the voxel location $i$, and $s$ is the smoothness factor. The segmentation loss is a binary cross-entropy loss with label smoothing \cite{Szeg16}. We apply label smoothing to encourage the network to be less confident in the target pseudo labels. The joint emphysema probability map is then generated by clamping the values of the summation of two dense regression activation maps between 0 and 1. The final loss $L$ for regression training is the sum of the interval regression loss, the overlapping loss, and the segmentation loss as $L = L_{INT} + L_{OL} + L_{SEG}$.

We used a variety of data augmentation techniques in training both networks, including flips and rotations, intensity and contrast jittering, cropping, Gaussian smoothing, and additive noise. These augmentations were applied randomly and all spatial transforms were preserved using trilinear interpolation to maintain the original size of the images.

Both networks were trained for a maximum of 200 epochs with an initial learning rate of 1e-5, using an Adam optimizer and a learning rate scheduler with exponential decay of 0.9. Training was terminate if there was no improvement in the validation set performance for 10 consecutive epochs.

\subsubsection*{Evaluation Metrics and Statistical Analysis}
To assess the classification performance, we use classification accuracy (ACC), the F-measurement (harmonic mean of precision and recall), and linear weighted kappa ($K$ statistics) as evaluation metrics. The ACC and F-measurement are calculated using the Sklearn python package (version 1.1.2) \cite{Pedr11}, while the $K$ statistics are calculated using the $rel$ software package in R (version 3.6.2). We also create confusion matrices and compute per-class metrics for each method. The level of agreement between algorithm predictions and visual scores is classified as slight, fair, moderate, good, or excellent based on $K$ values of 0.20 or less, 0.21–0.40, 0.41–0.60, 0.61–0.80, and 0.81 or higher, respectively \cite{Kund03}.

\section*{Results}
\begin{table}[ht]
\caption{Results of the classification and regression networks on the evaluation set ($n=7143$). The classification accuracy (ACC), F-measurements (F-measure), and linear weighted kappa are calculated against the visual scores. We also list the results from the Humphries algorithm, obtained on the same test set, where only ACC and kappa were reported for predicting centrilobular emphysema severity scores.}
\centering
\begin{tabular}{ |l|l|l|l|l|}
\hline
Method & Subtype & ACC(\%) & F-measure & Linear Weighted Kappa (95\% CI) \\
\hline
The Humphries algorithm \cite{Hump20} & centrilobular & 45 & - & 60 \\ \hline
Ours (classification)& centrilobular & 52.23 & 51.00 & 64.29 (63.16-65.42) \\ \hline
Ours (classification)& paraseptal & 59.12 & 57.12 & 42.03 (40.21-43.85) \\ \hline
Ours (regression)& centrilobular & 51.32 & 49.61 & 64.24 (63.14-65.35) \\\hline
Ours (regression)& paraseptal & 64.62 & 60.74 & 52.06 (50.40-53.73) \\ \hline
\end{tabular}
\label{tab:overall_results}
\end{table}

\begin{table}
  \caption{Confusion metrics of the classification and regression network for classifying centrilobular and paraseptal emphysema severity scores against the visual scores on the evaluation set ($n=7143$). Advanced destructive emphysema is denoted as Advanced.}
  \begin{subtable}[t]{0.9\textwidth}
  \centering
  \begin{tabular}[t]{l|llllll|l}
    \toprule
         &   & & &Visual Scores & &      &  \\\hline
    Predict     &  Absent    & Trace & Mild & Moderate & Confluence & Advanced & Precision (\%)  \\\hline
    Absent     & 1621 &  530 &  259&  21&  0&  0 &  66.68    \\
    Trace     & 625 &  451 &  322&  50&  0&  0  &  31.15  \\
    Mild     & 230 &  294 &  664&  288&  4&  0 &  44.86   \\
    Moderate     & 22 &  42 &  142&  514&  141&  7  &  59.22   \\
    Confluence     & 0 &  3 &  15&  153&  322&  42 &  60.19    \\
    Advanced  & 1 &  2 &  7&  23&  189&  159 &  41.73\\
    \bottomrule
     Recall (\%)    & 64.87 &  34.11 &  47.13&  49.00&  49.09&  76.44   &  52.23 (ACC\%)  \\
    \bottomrule
 \end{tabular}
\caption{Classification Network in Predicting Centrilobular Emphysema Severity}
\end{subtable}
\begin{subtable}[t]{0.9\textwidth}
\centering
 \begin{tabular}[t]{l|lll|l}
 \toprule
          &   & Visual Scores &   \\\hline
  Predict  &  Absent & Mild & Substantial & Precision (\%) \\\hline
  Absent     & 2508 &  740 &  181&  73.14     \\
  Mild   & 1144 &  942 &  467&  36.90    \\
  Substantial    & 205 &  183&  773&  66.58     \\
  \bottomrule
Recall (\%) & 65.02 & 50.51&  54.40&  59.12 (ACC\%)\\
  \bottomrule
\end{tabular}
\caption{Classification Network in Predicting Paraseptal Emphysema Severity}
\end{subtable}
\newline
\vspace*{0.5 cm}
\newline
 \begin{subtable}[t]{0.9\textwidth}
  \centering
  \begin{tabular}[t]{l|llllll|l}
    \toprule
         &   & & &Visual Scores & &      &  \\\hline
    Predict    &  Absent    & Trace & Mild & Moderate & Confluence & Advanced & Precision (\%)  \\\hline
    Absent     & 1734 &  536 &  206&  17&  0&  0 &  69.55    \\
    Trace     & 696 &  635 &  682&  142&  2&  0  &  29.43  \\
    Mild     & 56 &  121 &  393&  319&  21&  0 &  43.19  \\
    Moderate     & 11 &  26 &  117&  453&  192&  8  &  56.13   \\
    Confluence     & 2 &  3 &  10&  112&  334&  83 &  61.09    \\
    Advanced  & 0 &  1 &  1&  5&  107&  117 &  50.65\\
    \bottomrule
     Recall (\%)     & 69.39 &  48.03 &  27.89&  43.18&  50.91&  56.25   &  51.32 (ACC\%)  \\
    \bottomrule
\end{tabular}
\caption{Regression Network in Predicting Centrilobular Emphysema Severity}
\end{subtable}

\begin{subtable}[t]{0.9\textwidth}
\centering
\begin{tabular}[t]{l|lll|l}
 \toprule
          &   & Visual Scores &   \\\hline
  Predict  &  Absent & Mild & Substantial & Precision (\%) \\\hline
  Absent     & 2976 &  839 &  104&  75.94     \\
  Mild   & 784 &  778 &  455&  38.57    \\
  Substantial    & 97 &  248&  862&  71.42     \\
  \bottomrule
Recall (\%) & 77.16 & 41.72&  60.66&  64.62 (ACC\%)\\
  \bottomrule
\end{tabular}
\caption{Regression Network in Predicting Paraseptal Emphysema Severity}
\end{subtable}
\label{tab:cm}
\end{table}
We compared the performance of our classification and regression networks in predicting both centrilobular and paraseptal visual scores on an evaluation set. We also compared our results to those obtained using the Humphries algorithm for classifying centrilobular severity scores on the same evaluation set. As shown in Table~\ref{tab:overall_results}, both the classification and regression networks outperformed the Humphries algorithm in terms of overall classification accuracy for predicting centrilobular severity scores. Both networks achieved predictive accuracy above 51\%, while the Humphries algorithm only reached 45\%. In terms of kappa statistics, the agreements between automated scores and visual scores for both the classification and regression networks were good for centrilobular emphysema, slightly better than the moderate agreement reported by the Humphries algorithm. The agreements between our networks and paraseptal visual scores were moderate. The Humphries algorithm did not report results for paraseptal emphysema. Interestingly, the comparison between the classification and regression networks showed that the classification network outperformed the regression network in predicting centrilobular severity scores, but underperformed when predicting paraseptal scores in terms of overall accuracy and kappa statistics. These findings suggest that the two models may fit differently into the underlying target distribution.

Examining the confusion matrices in Table~\ref{tab:cm} (a,b), we observed that the classification network performed poorly in recognizing trace levels of centrilobular emphysema, with 31.15\% precision and 34.11\% recall. Analysis of the results revealed that most of these errors were caused by mislabeling trace emphysema as absent or vice versa. Similarly, in distinguishing paraseptal severity scores, the classification network had a high error rate for mild emphysema, with 36.90\% precision and 50.51\% recall. Furthermore, the regression network also exhibited a high rate of mislabeling between absence, trace and mild centrilobular emphysema. Specifically, the regression network performed the worst when labeling trace and mild centrilobular emphysema, with 29.43\% precision and 48.03\% recall in trace, and 43.19\% precision and 27.89\% recall in mild. In distinguishing paraseptal severity scores, the regression network also had the worst performance for mild emphysema, with 38.57\% precision and 41.72\% recall. The suboptimal performance observed in the lighter grades of emphysema severity may be attributed to the nuanced distinctions in lung involvement that define these grades (scores less than 2), resulting in a high degree of ambiguity in the scoring process. This challenge is also reflected in the results of the Humphries algorithm (Table~\ref{tab:cm_baseline}), highlighting the prevalent difficulty in accurately assessing the severity of emphysema. We identified a discrepancy in the performance of the Humphries algorithm and our methods in classifying the severity of centrilobular emphysema. Specifically, the Humphries algorithm demonstrated a tendency to overestimate the presence of emphysema in cases where it is actually absent, with 1495 cases being mislabeled as trace. Conversely, our methods tended to underestimate the severity in cases where it is present in minimal levels (trace), resulting in 536 cases of trace emphysema being mislabeled as absence by the regression network.

\begin{table}
  \caption{Confusion matrix of the Humphries algorithm for classifying centrilobular emphysema severity scores against the visual scores on the evaluation set ($n=7143$). Advanced destructive emphysema is denoted as Advanced.}
  \begin{subtable}[t]{0.9\textwidth}
  \centering
  \begin{tabular}[t]{l|llllll|l}
    \toprule
         &   & & &Visual Scores & &      &  \\\hline
    Predict     &  Absent    & Trace & Mild & Moderate & Confluence & Advanced & Precision (\%)  \\\hline
    Absent     & 637 &  126 &  35&  2 &  0&  0 &  79.62    \\
    Trace     & 1495 &  751 &  380&  23&  1&  0  &  28.82  \\
    Mild     & 324 &  377 &  678 &  166&  4&  0 &  43.77   \\
    Moderate     & 41 &  66 &  296 &  643 &  154&  8  &  53.22   \\
    Confluence     & 2 &  2 &  20&  211&  428&  108 &  55.51    \\
    Advanced  & 0 &  0 &  0&  4&  69&  92 &  55.57\\
    \bottomrule
     Recall (\%)    & 25.49 &  56.80 &  48.11&  61.29&  65.24&  44.23   &  45.21 (ACC\%)  \\
    \bottomrule
 \end{tabular}
\end{subtable}
\label{tab:cm_baseline}
\end{table}

The regression network has an advantage in terms of error distribution in confusion matrices, as it produces a percentage of lung involvement by averaging the dense regression maps within the lung volume. This improves its consistency in predicting severity scores. We observed a consistent shift in the error distribution in the confusion matrix, with the network more often mislabeling the target as one grade lower (Table~\ref{tab:cm} (c,d)). For example, in the case of centrilobular classification, the regression network classified 536 scans as absent and 121 scans as mild emphysema when they were manually scored as trace emphysema, and mislabeled 682 scans as trace and 117 scans as moderate when the actual labels were mild emphysema. In contrast, the error distribution was not consistent in the results of the classification network, which made more errors on one grade higher in the confluent centrilobular emphysema, with 141 scans mislabeled as moderate and 189 scans as advanced emphysema. However, for other grades, errors were made by mislabeling them as one grade lower.

In the extreme cases, the classification network mislabeled one case of absent centrilobular emphysema as advanced emphysema and classified 205 cases as substantial paraseptal emphysema when the visual score indicated absent. These critical errors suggest that the classification network was sometimes confused between centrilobular and paraseptal CT patterns, possibly due to the co-existence of both subtypes in some training scans. We also noticed that the regression network produced fewer critical errors in paraseptal classification, mistakenly labeling 97 scans as substantial when they were manually scored as absent. This number was 205 in the result of the classification network. This improvement may be due to the introduction of training constraints in the regression training, which enforce that dense predictions for centrilobular and paraseptal emphysema are mutually exclusive. However, we also observed that the regression network labeled 104 scans as absent when the visual scores indicated substantial paraseptal emphysema. This suggests that the regression network may under-segment or completely miss paraseptal emphysema in some cases.

\subsection*{Visual Interpretation}

\begin{figure*}
\minipage{1.0\textwidth}
\caption*{Case I: Confluence in centrilobular, Absent in Paraseptal Emphysema.}
\minipage{0.49\textwidth}
  \includegraphics[width=\linewidth]{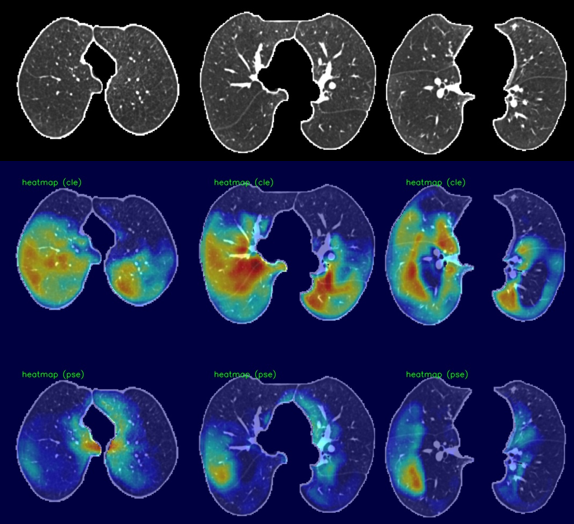}
  \caption*{(a) Class Activation Maps. Predicted Advanced for centrilobular, Substantial for Paraseptal.}
\endminipage\hfill
\minipage{0.49\textwidth}
  \includegraphics[width=\linewidth]{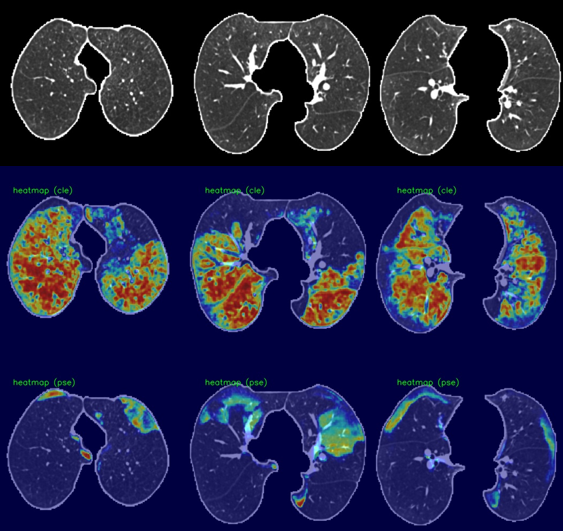}
  \caption*{(b) Regression Activation Maps. Predicted Advanced for centrilobular, Mild for Paraseptal.}
\endminipage\hfill
\endminipage\vfill
\minipage{1.0\textwidth}
\caption*{Case II: Advanced in centrilobular, Substantial in Paraseptal Emphysema.}
\minipage{0.49\textwidth}
  \includegraphics[width=\linewidth]{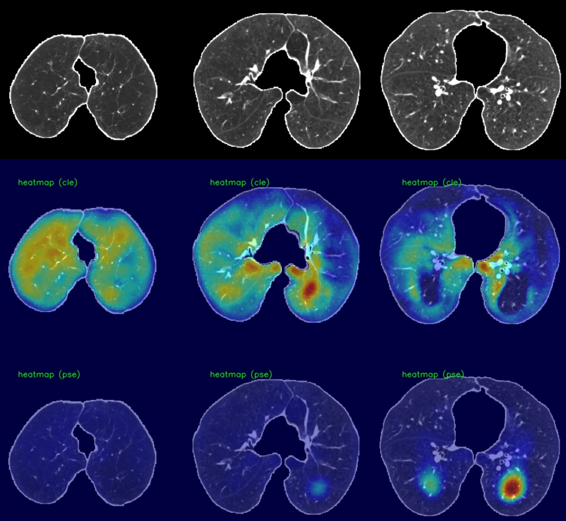}
  \caption*{(a) Class Activation Maps. Predicted Advanced for centrilobular, Absent for Paraseptal.}
\endminipage\hfill
\minipage{0.49\textwidth}
  \includegraphics[width=\linewidth]{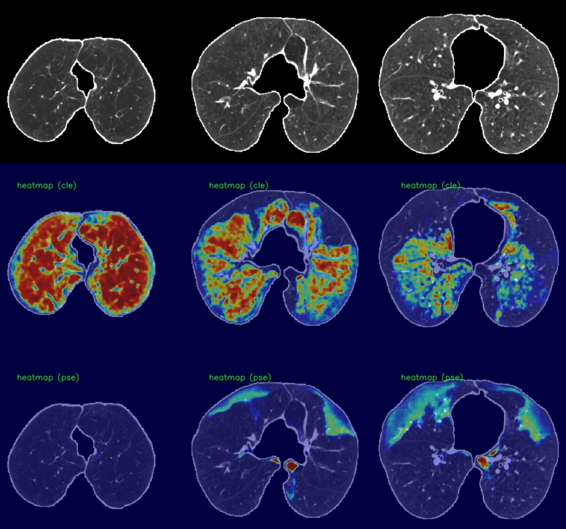}
  \caption*{(b) Regression Activation Maps. Predicted Advanced for centrilobular, Mild for Paraseptal.}
\endminipage\hfill
\endminipage\vfill

\caption{Dense class activation maps (left) versus dense regression activation maps (right), We show two cases, and each consists of three rows. The first row shows the input image (cropped and masked by the lung segmentation), the second row illustrates the activation maps for the centrilobular emphysema, and the third row shows the activation maps for the paraseptal emphysema.}
\label{fig:vis}
\end{figure*}

We utilized dense activation maps to visualize the features that correspond to the classification decisions. As shown in Figure~\ref{fig:vis}, we present the dense class and regression activation maps on two examples of predicting centrilobular and paraseptal emphysema subtypes. Each example consists of two images, with $3\times 3$ tiles, where the left image displays the dense class activation maps and the right image displays the dense regression activation maps. The three columns in each image are sampled axial slices from the input CT scan. For each image, the rows represent the preprocessed input CT scan, the activation map for centrilobular emphysema, and the activation map corresponding to paraseptal emphysema.

In general, the dense class activation maps do not necessarily align with object contours, e.g., blobs of paraseptal emphysema in the second case (Fig.~\ref{fig:vis} Case II (a)). Naturally, class activation maps only reflect discriminative regions responsible for classification. By utilizing the reconstruction network to generate dense features, our network's dense class activation maps already provide improved localization compared to the class activation maps generated by the Fleischer algorithm (which tend to be blurry blobs, as seen in their publication). The application of regression training further improves lesion localization, as can be seen in the subpleural paraseptal emphysema in the second case (Figure~\ref{fig:vis}, Case II(b)) and small blobs following the secondary lobular structures in the first case (Figure~\ref{fig:vis}, Case I(b)). Additionally, due to the use of the overlapping loss (Eq.\ref{eq:overlapping_loss}), the centrilobular and paraseptal activations do not overlap in the dense regression activation maps (Figure\ref{fig:vis}, Case II(b)), unlike in the class activation maps. For instance, in the first case, both class activation maps responded to the same regions in the right lobe (Figure~\ref{fig:vis}, Case I(a)). This highlights the effectiveness of our proposed method in providing improved lesion localization compared to the Fleischer algorithm.

\subsection*{Ablation Study On Neural Network Architecture}
\setreviewsoff
In our ablation study, we assessed three variants of ResNet architectures as the backbone network. ResNet18 comprised 2, 2, 2, and 2 ResNet layers. \add{Both} ResNet34 and ResNet50 consisted of 3, 4, 6, and 3 ResNet layers. \add{ResNet50 uses BottleNeck blocks, while ResNet34 uses the basic ResNet blocks \cite{He16}}. As reported in Table~\ref{tab:ablation}, ResNet34 has the largest number of parameters (64.79 million) compared with the ResNet18 (34.48 million) and ResNet50 (47.86 million). ResNet50 has more convolution operations than ResNet34 but has fewer number of parameters because of replacing ResNet blocks (two $3 \times 3 \times 3$ convolutions) in ResNet34 with BottleNeck blocks (one $3 \times 3 \times 3$ convolutions connecting two $1 \times 1 \times 1$ convolutions). \add{BottleNeck blocks employed in the ResNet50-based networks greatly reduce the number of parameters in 3D convolutions. However, the use of BottleNeck blocks in the ResNet50-based network leads to convolution features at the lowest resolution having a significantly larger number of channels (2048) compared to the ResNet34-based network (512)  . As a consequence, there is a substantial increase in the number of GMACs in the first upsampling layer, thereby contributing to an overall increase in GMACs for the ResNet50-based network when compared to the ResNet34-based network}. Details regarding the BottleNeck blocks can be found elsewhere \cite{He16}. Among the tested ResNet architectures, ResNet50 exhibited the highest computational intensity, reaching approximately 1702 GMACs (Giga Multiply-Add Operations). 

ResNet34 demonstrates superior predictive accuracy compared to ResNet18 and ResNet50 in both the classification and regression networks. This suggests that volumetric data may necessitate higher complexity (more parameters) for optimal fitting. Notably, the regression networks generally attained higher performance (especially in linear weighted Kappa values) than the classification networks in predicting paraseptal labels. This improvement could be attributed to the incorporation of mutual exclusive constraints in regression training, as discussed in Section \ref{eq:overlapping_loss}. ResNet18 exhibits the lowest performance, indicating that the smaller network's complexity may not be adequate for effectively addressing the subtyping problem. We also observed that the linear-weighted kappa values are generally higher than accuracy and F-measure in centrilobular predictions, while kappa values are lower than accuracy and F-measure in paraseptal predictions. These disparities are particularly pronounced in the results of the classification networks. This indicates a heightened occurrence of critical errors when attempting to predict paraseptal labels, as also evidenced by the observed confusion matrices (refer to Table~\ref{tab:cm}) in the results of ResNet34. 

\begin{table}
\caption{Results of Ablation Study of three variants (ResNet18, ResNet34, and ResNet50) of ResNet as our backbone network.} 
\begin{subtable}[t]{0.9\textwidth}
\centering
\begin{tabular}[t]{l|llll|ll}
\toprule
Backbone    &  Subtype    & Accuracy (\%) & F-measure & Linear Weighted Kappa (95\% CI) & \#Param (M) & GMACs \\\hline
\multirow{2}{*}{ResNet18}&Centrilobular &48.55 &  47.58 &  62.50 (61.83-63.63) &  \multirow{2}{*}{34.48} & \multirow{2}{*}{1148.05}  \\
                         & Paraseptal & 54.81 &  51.68&  33.63 (31.77-35.49) &   &   \\\hline
\multirow{2}{*}{ResNet34}&Centrilobular&52.33 &  51.00&  64.29 (63.16-65.42)&  \multirow{2}{*}{64.79}&  \multirow{2}{*}{1661.95}   \\
                        & Paraseptal& 59.12 & 57.12&  42.03 (40.21-43.85)&  &      \\\hline
\multirow{2}{*}{ResNet50}&Centrilobular&51.31 &  51.00 &  63.00 (61.84-64.16) &  \multirow{2}{*}{47.86}&  \multirow{2}{*}{1702.69}  \\
                        &Paraseptal &54.79 &  52.22 &  34.62 (32.76-36.48) &  &     \\\hline
\bottomrule
\end{tabular}
\caption{Classification Network in Predicting Paraseptal Emphysema Severity}
\end{subtable}

\begin{subtable}[t]{0.9\textwidth}
\centering
\begin{tabular}[t]{l|llll|ll}
\toprule
Backbone    &  Subtype    & Accuracy (\%) & F-measure & Linear Weighted Kappa (95\% CI) & \#Param (M) & GMAC \\\hline
\multirow{2}{*}{ResNet18}&Centrilobular &49.91 &  47.08 &  62.53 (61.39-63.67) &  \multirow{2}{*}{34.48} & \multirow{2}{*}{1147.81}  \\
                         & Paraseptal & 55.87 &  53.92&  39.16 (37.40-40.92) &   &   \\\hline
\multirow{2}{*}{ResNet34}&Centrilobular&51.32 &  49.61&  64.24 (63.14-65.35)&  \multirow{2}{*}{64.79}&  \multirow{2}{*}{1661.72}   \\
                        & Paraseptal& 64.62 & 60.74&  52.06 (50.40-53.73)&  &      \\\hline
\multirow{2}{*}{ResNet50}&Centrilobular&51.03 &  46.70 &  62.15 (60.98-63.31) &  \multirow{2}{*}{47.86}&  \multirow{2}{*}{1702.45}  \\
                        &Paraseptal &62.21 &  56.22 &  47.21 (45.45-48.97) &  &     \\\hline
\bottomrule
\end{tabular}
\caption{Regression Network in Predicting Emphysema Severity}
\end{subtable}

\label{tab:ablation}
\end{table}

\section*{Discussion}
In this study, we propose a novel approach for automating the Fleischer scoring system for identifying emphysema subtypes on CT images using deep neural networks. Our method incorporates a reconstruction sub-network, which is utilized to generate high-resolution activation maps, providing more localized information for model interpretation in comparison to the low-resolution heatmaps produced by the Fleischer algorithm. Additionally, our regression-based training strategy offers an estimation of emphysema percentage in the lungs and generates the regression activation maps which offer improved emphysema localization. The regression training approach is more intuitive as it utilizes the semantic link between severity scores and emphysema involvement in the lungs, compared to the classification approach which only aims to identify discriminative features between different severity grades.

The target labels for regression training are percentage intervals translated from categorical scores using a predefined mapping table, and the network is trained using an interval regression loss to ensure that the predicted percentage falls within the target interval. This approach could be used for automating many other scoring systems used in radiology based on visual assessment of the (relative) size of the affected volume. We use lung-wise average pooling to aggregate the sigmoid-activated dense features (dense regression activation maps) to estimate the emphysema percentage per lung, and we incorporate the overlapping loss to ensure that each voxel is only assigned to one of the two emphysema subtypes (centrilobular or paraseptal). Additionally, we use low-attenuation areas in the lung (LAA-950) as visual cues in the regression training to provide localized information to the network.

Our method generates both categorical visual scores and estimated emphysema percentages for both centrilobular and paraseptal subtypes, offering additional features in comparison to the existing method, which can only produce centrilobular severity scores. The dense regression activation maps generated by our approach provide detailed emphysema localization, potentially enabling further clinical research in this field.

The results of our study showed that our method (both the classification and regression networks) outperformed the Fleischer algorithm in terms of classification accuracy for predicting centrilobular severity grades (52\% versus 45\%). Our method also had better reader agreement as measured by kappa statistics.

The regression approach has several advantages compared to the classification approaches. It resulted in fewer critical errors in the confusion matrix comparison, such as mislabeling heavily diseased cases as disease-free. In addition, the errors were distributed more consistently, with a shift to a lower severity grade in the confusion matrix (Table~\ref{tab:cm} (c)), which is not observed in the results of the classification approach (Table~\ref{tab:cm} (a)).

The ablation study (Refer to Table~\ref{tab:ablation}) showcased consistent observations regarding the classification performances and their patterns across various subtyping tasks using different ResNet backbones.

In terms of qualitative analysis, the dense regression activation maps were observed to provide superior localization, particularly in the case of paraseptal emphysema with sub-pleural bullae. In contrast, the dense class activation maps were observed to be less specific, as they included many surrounding regions, and were not able to distinguish different emphysema subtypes. 

There are several limitations to consider in this study. First, our systems were only trained using CT scans from the COPDGene study, which had a specific data acquisition protocol and carefully curated scans. This may limit the generalizability of our algorithms to other datasets with different CT acquisition processes. We attempted to mitigate this risk by using data augmentation techniques that introduce common noise patterns and using early stopping to prevent overfitting. Second, our algorithms require the availability of lung segmentation for both training and inference, which may be a challenge to implement in clinical practice, although fast and accurate publicly available systems for CT lung segmentation are available \cite{Hofm20,Xie21}. Third, we did not validate the segmentation performance using the generated dense regression activation maps, although these maps appear to be capable of localizing emphysema patterns based on visual inspection. This was due to the lack of voxel-wise annotations of emphysema patterns with different subtypes.

\section*{Model Availability}
Our algorithms are available at GitHub (\url{https://github.com/DIAGNijmegen/bodyct-dram-emph-subtype}). We integrated our trained regression network as a ready-to-use web service hosted on the Grand-challenge platform (\url{https://grand-challenge.org/algorithms/weakly-supervised-emphysema-subtyping/}). 

\section*{Data Availability}
To access the COPDGene data used in this study for research purposes, please visit \url{https://www.ncbi.nlm.nih.gov/projects/gap/cgi-bin/study.cgi?study_id=phs000179.v6.p2} and submit an ancillary study proposal. We received approval for this work under the ANC-251 proposal. To submit your proposal, contact the COPDGene Administrative Core Executive Secretary, Sara Penchev, at PenchevS@NJHealth.org. The corresponding author (bram.vanginneken@radboudumc.nl) may be reached for data inquiry.

\bibliography{fullstrings,medlinestrings,reference}
\section*{Competing interests}
B.v.G. is founder and shareholder of Thirona. The other authors declare no competing interests.

\section*{Author Contribution Statement}
B.v.G., D.J.S., and W.X. conceptualized and designed the project. J.P.C. and W.X. performed the data collection. B.v.G., C.J. and J.P.C. supervised and coordinated the whole work. W.X. created the computational codes and carried out all the experiments. W.X. provided the graphs and drafted the manuscript. B.v.G. participated in writing and editing the manuscript. B.v.G. and D.J.S. reviewed and contributed to revising the final manuscript.

\section*{Acknowledgements}
The Dutch Lung Foundation, under project 5.1.17.171 supported this work. In addition, we acknowledge the COPDGene Study (ancillary study ANC-251) for providing the data used. COPDGene is funded by Award Number U01 HL089897 and Award Number U01 HL089856 from the National Heart, Lung, and Blood Institute. The COPD Foundation also supports the COPDGene study (NCT00608764) through contributions to an Industry Advisory Committee comprised of AstraZeneca, Bayer Pharmaceuticals, Boehringer-Ingelheim, Genentech, GlaxoSmithKline, Novartis, Pfizer, and Sunovion.

We especially thank Dr. Stephen M. Humphries from National Jewish Health for sharing the data selection and partitioning in their study, which is used as a comparison in this paper.
\end{document}